\documentclass{revtex4}
\usepackage{epsf}

\begin{document}
\title{Gamow-Teller transitions from $^{9,11}$Li to $^{9,11}$Be}

\author{Yoshiko Kanada-En'yo}
\affiliation{Yukawa Institute for Theoretical Physics, Kyoto University,
Kyoto 606-8502, Japan}

\begin{abstract}
Gamow-Teller(GT) transitions in the $\beta$ decays of $^9$Li and $^{11}$Li are investigated 
with theoretical calculations of antisymmetrized molecular dynamics.
The calculated $B(GT)$ values are small for transitions to low-lying states in Be isotopes,  
while relatively large $B(GT)$ values are found for excited states at excitation energy 
$E_x\ge 10$ MeV.
Sum of the $B(GT)$ values are discussed for each spin parity of final states.
The calculated results seem to be inconsistent with the experimental report
of the strongest GT-transition from $^9$Li to the $5/2^-$ state at 11.8 MeV of $^9$Be.
\end{abstract}

\maketitle

\noindent

\section{Introduction}
Experimental studies of $\beta$ decays from unstable nuclei near 
the drip lines are developing nowaday due to recent progress of experimental techniques.
Gamow-Teller(GT) transition strength $B(GT)$ values were extracted not only for transitions 
to low-lying states but also those to highly excited states.
Moreover, determination of the $B(GT)$ values has been performed also 
in high resolution experiments of charge exchange reactions as well as the $\beta$ decays.

In these years, measurements of the $\beta$ decays of $^9$Li and $^{11}$Li were carried out in several
experiments, for example, at the ISOLDE facility in CERN.
They provide information of new states in $^9$Be and $^{11}$Be
such as the excitation energies, spins and $B(GT)$ values.
In the $\beta$ decay of $^9$Li, the GT transitions to low-lying $^9$Be states in $E_x\le 10$ MeV
are weak while the strong transitions to the 11.81 MeV state were reported\cite{Nyman:1989pj,Chou:1993zz}. 
The extracted $B(GT)$ value for the $^9$Be state at $E_x=11.81$ MeV is surprizingly large as
$B(GT)=8.9(1.9)$ which takes 65 \% of the Ikeda sum rule, $9 (g_A/g_V)^2\sim 13.4$.
In a new measurement of the $\beta$ decay\cite{Prezado03}, 
its spin and parity were assigned to be $5/2^-$. 
Compared with the mirror transitions from $^9$C to 
the $^9$B($5/2^-$) state at 12.19 MeV\cite{Bergmann01}, 
this suggests the abnormally large mirror asymmetry that 
the $B(GT)$ value for the $^9$Be($5/2^-$) at 11.81 MeV is
larger by factor 4 than that for the mirror transition as argued in Ref.~\cite{Prezado03}.
However, the shell-model calculations in a $p$-shell basis do not indicate such the large $B(GT)$ values 
in the $^9$Li decays nor the large mirror asymmetry\cite{Millener05}.

The experimental measurements of the $\beta$ decay were performed also 
for the drip-line nucleus, $^{11}$Li\cite{Aoi97,Fynbo04,Hirayama05,Madurga:2009ce}.
Many excited states of $^{11}$Be were observed in low energy region.
The GT transitions to $^{11}$Be states in $E_x\le 10$ MeV are not strong and
the extracted $B(GT)$ values are less than $B(GT)=0.5$ for the low-lying states\cite{Hirayama05}.
In a recent measurement, a new state at 18 MeV in $^{11}$Be was observed
and the value $B(GT)=1.9$ for this state was extracted from the branching 
ratios\cite{Madurga:2009ce}.

As mentioned above, in the $\beta$ decays of neutron-rich Li isotopes, the GT transitions 
to low-lying states of $^9$Be and $^{11}$Be are relatively weak 
compared with rather strong GT transitions to highly excited states.
It is understood easily because of the cluster features of Be isotopes.
In theoretical studies of Be isotopes, it was suggested that 
$2\alpha+n$ and $2\alpha+3n$ cluster structures develop in 
most of the low-lying states of $^9$Be and $^{11}$Be, respectively
\cite{Zhan76,Okabe77,SEYA,OERTZEN,Arai:1996dq,Enyo-be11,Oertzen-rev}.
The GT transitions to such the cluster states should be suppressed
because GT transition to an ideal $\alpha$ cluster written by the $(0s)^4$ configuration
is forbidden exactly due to Pauli principle.
In other words, the high-lying states in $E_x \ge 10$ MeV 
with the strong GT transitions are expected to have significant components of cluster breaking.

In this paper, we investigate the GT transition strengths of the $\beta$ decays, 
$^{9}$Li$\rightarrow ^{9}$Be and $^{11}$Li$\rightarrow ^{11}$Be, with a theoretical method of 
antisymmetrized molecular dynamics(AMD)\cite{ENYObc,ENYOsup,AMDrev}.
Distribution of the $B(GT)$ values is analyzed for each spin of the final states. 
Particular attention is paid to the strong GT transitions to the $^9$Be $5/2-$ state at 11.81 MeV
suggested in the experimental report.

This paper is organized as follows. In the next section, we describe the formulation of the present calculations.
We explain the adopted effective interactions in \ref{sec:interactions}, and 
show the calculated results as well as the experimental data in \ref{sec:results}.
Finally, in \ref{sec:summary}, we give a summary and an outlook.

\section{Formulation} \label{sec:formulation}

We use a method of AMD which has been proved to be one of the useful approaches for 
structure study of light unstable nuclei. In the present work, 
we first perform the variation after projection(VAP) with respect to 
the total-angular-momentum and parity projection(spin-parity projection)
in the AMD~\cite{Enyo-c12} and extend the method
to calculate GT strength functions to final states in a wide energy region.
The detailed formulation of the basic AMD method 
for nuclear structure study is described in Refs.~\cite{ENYObc,ENYOsup,AMDrev}.

In the AMD method, a wave function of a $A$-nucleon system is 
written by a Slater determinant of single-particle 
Gaussian wave packets, 
\begin{equation}
 \Phi_{\rm AMD}({\bf Z}) = \frac{1}{\sqrt{A!}} {{\cal A}} \{
  \varphi_1,\varphi_2,...,\varphi_A \},
\end{equation}
where the $i$th single-particle wave function is written by a product of
spatial($\phi$), intrinsic spin($\chi$) and isospin($\tau$) 
wave functions,
\begin{eqnarray}
 \varphi_i&=& \phi_{{\bf X}_i}\chi_i\tau_i,\\
 \phi_{{\bf X}_i}({\bf r}_j) &\propto& 
\exp\bigl\{-\nu({\bf r}_j-\frac{{\bf X}_i}{\sqrt{\nu}})^2\bigr\},
\label{eq:spatial}\\
 \chi_i &=& (\frac{1}{2}+\xi_i)\chi_{\uparrow}
 + (\frac{1}{2}-\xi_i)\chi_{\downarrow}.
\end{eqnarray}
$\phi$ and $\chi$ are represented by 
complex variational parameters, ${\rm X}_{1i}$, ${\rm X}_{2i}$, 
${\rm X}_{3i}$, and $\xi_{i}$. The isospin
function $\tau_i$ is fixed to be up(proton) or down(neutron). 
We take a fixed width parameter $\nu$ which is optimized for each nucleus.
That is $\nu=0.20$ fm$^{-2}$ for $^9$Be and $^9$Li, and 
$\nu=0.18$ fm$^{-2}$ for $^{11}$Be and $^{11}$Li in the present calculations.
Accordingly, an AMD wave function
is expressed by a set of variational parameters, ${\bf Z}\equiv 
\{{\bf X}_1,{\bf X}_2,\cdots, {\bf X}_A,\xi_1,\xi_2,\cdots,\xi_A \}$.

For the lowest $J^\pi$ state,
we vary the parameters ${\bf X}_i$ and $\xi_{i}$($i=1\sim A$) to
minimize the energy expectation value of the Hamiltonian,
$\langle \Phi|H|\Phi\rangle/\langle \Phi|\Phi\rangle$,
with respect to the spin-parity projected AMD wave function;
$\Phi=P^{J\pi}_{MK}\Phi_{\rm AMD}({\bf Z})$.
Here, $P^{J\pi}_{MK}$ is the spin-parity projection operator.
After the energy variation, the optimized parameter set ${\bf Z}^{J\pi}_1$ of the minimum-energy solution
for the lowest $J^\pi$ state is obtained.
The solution ${\bf Z}^{J\pi}_n$ 
for the $n$th $J^\pi$ state is determined by varying ${\bf Z}$ 
so as to minimize the energy of the wave function orthogonalized to the lower states; 
\begin{equation}
|\Phi\rangle =|P^{J\pi}_{MK}\Phi_{\rm AMD}({\bf Z})\rangle
-\sum^{n-1}_{k=1}
|\Phi^{(k)J\pi}_{MK}\rangle 
\langle \Phi^{(k)J\pi}_{MK} 
|P^{J\pi}_{MK}\Phi_{\rm AMD}({\bf Z})\rangle,
\end{equation}
where $\Phi^{(k)J\pi}_{MK}$ is the normalized wave function determined for the lower states.
This is the standard procedure of the VAP calculation in the AMD method.

In the VAP calculations, the wave functions for the $J^\pi_n$ states 
are obtained one by one from the lower energy states and it is not easy to calculate 
all final states of the GT transitions which fragment generally into many high-lying states.
In order to exhaust the GT strengths from an initial state, we extend the basis wave functions 
by operating one-body spin-isospin operators to the VAP wave functions for the parent nucleus.
Let us consider the $\beta^-$ decay of $^{9}$Li.
We perform the VAP calculation of the lowest $J^\pi=3/2^-$ state of $^9$Li 
and obtain the wave function $\Phi^{^9{\rm Li}}_{\rm AMD}({\bf Z}^{3/2-}_1)$ for 
the $^{9}$Li ground state. Here we rewrite ${\bf Z}^{3/2-}_1$ for $^9$Li with ${\bf Z}_{\rm init}$.
Final states of the GT transitions from the $^{9}$Li ground state are
$J^\pi_n=1/2^-,3/2^-$ and $5/2^-$ states in the daughter nucleus $^9$Be.
As for the final states in $^9$Be, we prepare the basis wave functions in two steps as follows. 
We first do the VAP calculations for the lowest two $J^\pi$ states of $^9$Be, and  
obtain the wave functions $\Phi^{^9{\rm Be}}_{\rm AMD}({\bf Z}^{J\pi}_n)$ for 
$J^\pi_n=1/2^-_1$, $1/2^-_2$,  $3/2^-_1$,  $3/2^-_2$, $5/2^-_1$ and  $5/2^-_2$ 
which approximately describe the corresponding low-lying states.
Next, we create other basis wave functions by operating the one-body spin-isospin operators 
to the $\Phi^{^9{\rm Li}}_{\rm AMD}({\bf Z}_{\rm init})$ as,
\begin{equation}\label{eq:gt-single}
 \Phi_{k,\alpha}^{\sigma\tau}({\bf Z}_{\rm init}) = \frac{1}{\sqrt{A!}} {{\cal A}} \{
  \varphi_1,\varphi_2,\cdots,\sigma_\alpha\tau^{-}\varphi_k,\cdots,\varphi_A \},
\end{equation}
where $\varphi_k$ ($k=1,\cdots,6$) is a neutron wave function in $^9$Li.
$\sigma_\alpha\tau^{-}$ is the one-body GT transition operator 
where $\sigma_\alpha$($\alpha=x,y$ and $z$) is 
the spin operator and $\tau^{-}$ is the charge changing operator.
 As a result, we get $6\times 3=18$ number of 
basis wave functions, $\Phi_{k,\alpha}^{\sigma\tau}({\bf Z}_{\rm init})$, which exhaust the 
GT transition strengths from the $\Phi^{^9{\rm Li}}_{\rm AMD}({\bf Z}_{\rm init})$.

Finally we determine the wave functions of the final $^9$Be states 
by the spin-parity projection and the superposition of all the $6+18$ basis wave functions, 
\begin{equation}
\Phi_{Jn}=\sum_{J'Km} c^{(Jn)}_{J',K,m}
P^{J\pi}_{MK}\Phi^{^9{\rm Be}}_{\rm AMD}({\bf Z}^{J'\pi}_m)
+\sum_{Kk\alpha} c^{(Jn)}_{k,\alpha} P^{J\pi}_{MK}\Phi_{k,\alpha}^{\sigma\tau}({\bf Z}_{\rm init}).
\end{equation}
The coefficients $c^{(Jn)}_{J',K,m}$ and $c^{(Jn)}_{k,\alpha}$ are determined by diagonalizing Hamiltonian and
norm matrices. 

The value $B(GT)$ of the GT transition strength is written by the square of the reduced matrix element 
\begin{equation}
B(GT)=(g_A/g_V)^2|\langle {\cal O}^{(GT-)}_\mu \rangle|^2
\end{equation}
of the GT transition operator,
\begin{equation}
{\cal O}^{(GT\pm)}_\mu= \sum_i \sigma_\mu(i) \tau^\pm(i),
\end{equation}
where $g_A$ and $g_V$ are the axial-vector and vector coupling constants and taken to be
$(g_A/g_V)^2=1.51$ in the present calculations.
It is worth mentioning again that the GT transition strengths from the $^9$Li ground state written by the wave function
$P^{J\pi=3/2-}_{MK}\Phi^{^9{\rm Li}}_{\rm AMD}({\bf Z}_{\rm init})$ are exhausted by the final states, 
$\Phi_{Jn}$ ($n=1,\cdots,24$), of $^9$Be.

\section{Effective nuclear forces}\label{sec:interactions}

The effective nuclear interaction adopted in the present work
consists of the central force, the
spin-orbit force and the Coulomb force.
For the central force, the MV1 force case (3) \cite{TOHSAKI} containing 
finite-range two-body and zero-range three-body terms is used.
As for the spin-orbit force, the same two-range Gaussian form as 
that in the G3RS force \cite{LS} is adopted.

The used values of the interaction parameters in the MV1 force are 
$b=h=0$ and $m=0.62$ of Bartlett, Heisenberg and Majorana parameters, 
and the strengths of the spin-orbit force are taken to be 
$u_{I}=-u_{II}=3000$ MeV.
These parameters are the same as those used in the AMD+VAP calculations of
$^{12}$C and $^{10}$Be in Refs.~\cite{Enyo-c12,ENYO-be10}.
Hereafter, we call this parameter set (A).
In order to see the interaction dependence of the GT transition strengths, we also 
use two sets of modified parameters, (B) and (C).
The set (B) $b=h=0$, $m=0.62$ and $u_{I}=-u_{II}=2000$ MeV
has the same central force but the weaker spin-orbit force than (A), and the set (C) 
$b=h=0.15$, $m=0.62$ and $u_{I}=-u_{II}=3000$ MeV contains 
the Bartlett and Heisenberg terms in the central force but the same strengths of the spin-orbit force as (A).

\section{Results}\label{sec:results}

\subsection{GT transition from $^9$Li to $^9$Be}

In the present results of the VAP calculations for low-lying states of $^9$Be,
it is found that $2\alpha+n$ cluster structure developes in 
the ground state($3/2^-_1$) and excited states, $5/2^-_1$, $1/2^-_1$, $3/2^-_2$ and
$5/2^-_2$. The calculated energy spectra of these low-lying states are consistent with 
those calculated with the $2\alpha+n$ cluster models\cite{Zhan76,Okabe77,Arai:1996dq} and
the experimental energy levels(table \ref{tab:bgt-be9}).

\begin{figure}[th]
\epsfxsize=7 cm
\centerline{\epsffile{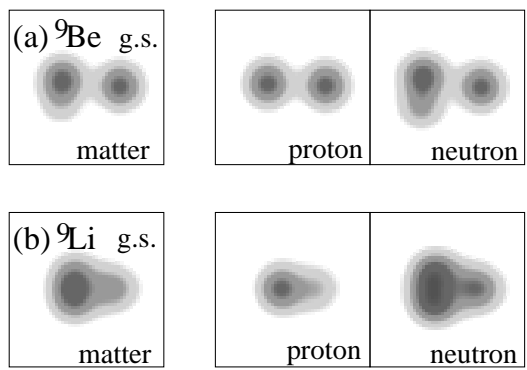}}
\vspace*{8pt}
\caption{\label{fig:be9-dense}
The distribution of matter, proton and neutron density in the intrinsic wave 
functions of (a)$\Phi^{^9{\rm Be}}_{\rm AMD}({\bf Z}^{3/2-}_1)$ for the
$^9$Be ground state 
and (b)$\Phi^{^9{\rm Li}}_{\rm AMD}({\bf Z}_{\rm init})$ for the
$^9$Li ground state calculated with the interaction (A).
}
\end{figure}

By adding the
basis wave functions $\Phi_{k,\alpha}^{\sigma\tau}({\bf Z}_{\rm init})$ 
constructed from the spin-isospin operated $^9$Li wave functions
to the VAP wave functions  as explained in \ref{sec:formulation}, 
the final wave functions of $^9$Be are obtained and
the $B(GT)$ values for the transitions $^9$Li$\rightarrow^9$Be are calculated.
The calculated $B(GT)$ values to the final $^9$Be states up to $E_x\sim 13$ MeV 
are shown in table \ref{tab:bgt-be9}. The $B(GT)$ values to
the ground state($3/2^-_1$), the $5/2^-_1$, $1/2^-_1$, $3/2^-_2$ and
$5/2^-_2$ states are small. This is because those states of $^9$Be have
the $2\alpha+n$ cluster structure and they have 
small overlap with the initial state of $^9$Li having no developed cluster structure.
Fig.~\ref{fig:be9-dense} shows the density distribution in the intrinsic wave functions of the 
$^9$Be ground state($\Phi^{^9{\rm Be}}_{\rm AMD}({\bf Z}^{3/2-}_1)$)
 and that of the $^9$Li ground state($\Phi^{^9{\rm Li}}_{\rm AMD}({\bf Z}_{\rm init})$).
It shows that the intrinsic  structure changes drastically from the initial state to the final state.
Moreover, the GT transitions to such the cluster states with the $2\alpha$ core are suppressed
because of Pauli principle as mentioned before. In other words, 
the finite $B(GT)$ values account for the slight dissociation of 
the ideal $\alpha$ clusters in $^9$Be.

In contrast to the small $B(GT)$ values to the low-lying states,
the excited states at $E_x=12\sim 13$ MeV show significant GT strengths
which come from the non-$\alpha$-cluster states constructed from the 
spin-isospin operated $^9$Li wave functions given in Eq. \ref{eq:gt-single}.
The strongest GT transition is found for the $1/2^-_2$ state in the present results.
In the experimental measurements of the $\beta$ decay of $^9$Li, 
rather large values of $B(GT)$ were reported for the states 
around $E_x=12$ MeV\cite{Nyman:1989pj,Chou:1993zz,Prezado03}
as $B(GT)=1.44$ for the $E_x=11.28$ MeV state and
$B(GT)=8.9$ for the $E_x=11.81$ MeV state. The 
former state is suggested to be a $3/2^-$ state\cite{Langevin81} and it might correspond to the 
theoretical $3/2^-_3$ state in the present result. The calculated value of $B(GT)=0.94$ agrees to
the experimental value. The spin and parity of the latter $E_x=11.81$ MeV state was assigned as 
$J^\pi=5/2^-$ in the recent analysis of the $\beta^-$ decay of $^9$Li\cite{Prezado03}. 
Although this state seems to correspond energetically to the calculated $5/2^-_3$ state,
the theoretical $B(GT)$ value is much smaller than the experimental value of $B(GT)=8.9$ which exhausts 
65 \% of the Ikeda sum rule.
As shown later, there is no $5/2^-$ state having such the strong GT strength compatible to $B(GT)=8.9$
in the present calculation.
Also in the shell-model calculations, it is difficult to reproduce the 
extraordinary large $B(GT)$ value of the experimental data\cite{Millener05}.

\begin{figure}[th]
\epsfxsize=7 cm
\centerline{\epsffile{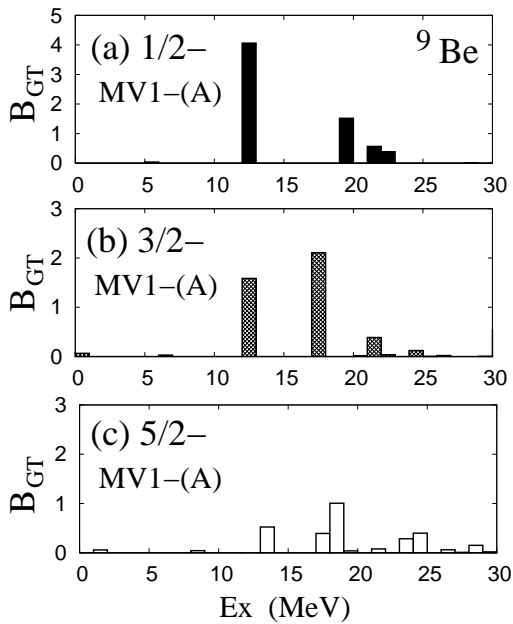}}
\vspace*{8pt}
\caption{The $B(GT)$ distribution to (a)$1/2^-$ states, 
(b)$3/2^-$ states and (c)$5/2^-$ states of $^9$Be in the $\beta$ decay of $^9$Li
calculated with the interaction (A). 
\label{fig:be9-gt}
}
\end{figure}

\begin{figure}[th]
\epsfxsize=7 cm
\centerline{\epsffile{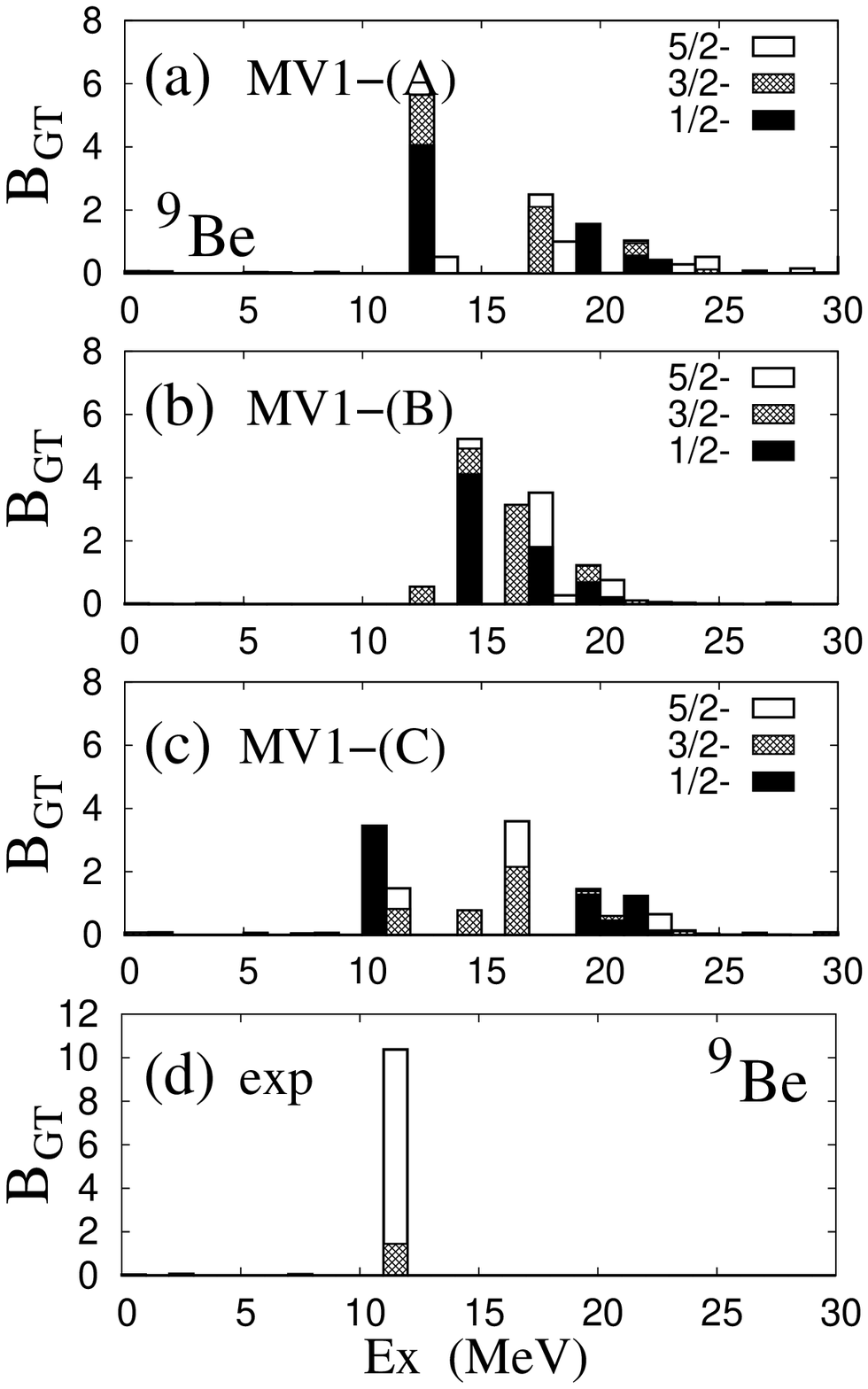}}
\vspace*{8pt}
\caption{
\label{fig:be9-gt-hist}
The $B(GT)$ distribution for the decay 
$^9$Li$\rightarrow ^9$Be. The $B(GT)$ values to $1/2^-$, $3/2^-$ and $5/2^-$ states
of $^9$Be are piled up.
The upper panels (a),(b) and (c) show 
the results calculated with the interaction sets, (A), (B) and (C), respectively. 
The lowest panel (d) shows the experimental data which are taken from the
same references as those in table \ref{tab:bgt-be9}.
}
\end{figure}

Figure \ref{fig:be9-gt} shows the $B(GT)$ distribution to excited states of
$^9$Be up to $E_x=30$ MeV calculated by using the interaction set (A). 
In the $B(GT)$ values to $1/2^-$ states, the largest peak
is found at $E_x\sim 12$ MeV, and some strengths distribute around 20 MeV.
The transition strengths to $3/2^-$ states distribute mainly in $E_x=10-20$ MeV region.
The $B(GT)$ values to $5/2^-$ states are relatively small compared with those to $1/2^-$ and $3/2^-$ states.
In order to show the interaction dependence of the GT strengtions, 
we show the $B(GT)$ distribution calculated with three interaction sets (A), (B) and (C) in 
Fig.~\ref{fig:be9-gt-hist}.
The $B(GT)$ distribution is qualitatively similar between these three interaction sets,
though the broadness of the distribution changes slightly.
In all the results, the $B(GT)$ values are very small for the transitions to low-energy 
$E_x < 10$ MeV states, while
the strengths distribute mostly in $E_x=10-20$ MeV.
Compared with the shell model calculations in Ref.~\cite{Suzuki:2003fw}, which show
the $B(GT)$ distribution concentrating at $E_x=10-12$ MeV, the present calculations show 
significant fraction of the $B(GT)$ values to highly excited states in $E_x > 15$ MeV.

\begin{table}[ht]
\caption{ 
\label{tab:bgt-be9} Experimental data of the GT strengths in the $\beta^-$ decays of $^9$C and $^9$Li,
and theoretical values of the $^9$Li decay. The experimental data of the B(GT) values
are taken from \cite{Chou:1993zz} and $^b$\cite{Bergmann01}. 
$^a$The spin and parity $3/2^-$ of the $^9$Li(11.28 MeV) is the assignment of Ref.~\cite{Langevin81}
but it is not confirmed yet and other spin-parity candidates, $7/2^-$ and $7/2^+$, are 
suggested\cite{Tilly04}.
$^c$$5/2^-$ was assigned in Ref.~\cite{Prezado03}.
} 
\begin{center}
\begin{tabular}{c|cccc|cc}
\hline
   & \multicolumn{2}{c}{exp.} & \multicolumn{2}{c}{exp.} & \multicolumn{2}{c}{cal.(A)}\\
   & $^9$B & $^9$C ($\beta^+$)    & $^9$Be & $^9$Li ($\beta^-$) & $^9$Be & $^9$Li ($\beta^-$) \\
$J^\pi$ & $E_x$  &  $B(GT)$ & $E_x$  &  $B(GT)$ & $E_x$  &  $B(GT)$ \\
\hline
$3/2^-_1$ & 0& 0.33 & 0 & 0.31 & 0 & 0.065 \\
$5/2^-_1$ & 2.361 & 0.021 & 2.43 & 0.054 & 2.0 & 0.057\\
$1/2^-_1$ & 2.80 & 0.015 & 2.78 & 0.011 & 5.3 & 0.038\\
$3/2^-_2$ &  & & 5.59 &  & 6.9 & 0.029\\
$5/2^-_2$ &  & & 7.94 & 0.048 & 8.2  & 0.043  \\
$1/2^-_2$ & & & & & 12.0 & 4.1 \\
$3/2^-_3$ & &  & 11.28$^a$ & 1.44 & 12.1 & 0.94 \\
$3/2^-_4$ & &  &  & & 12.4 & 0.64 \\
$5/2^-_3$ & 12.19$^b$ & 1.812$^b$ & 11.81$^c$ & 8.9 & 13.1 & 0.52 \\
\hline
\end{tabular}
\end{center}
\end{table}

\subsection{GT transition from $^{11}$Li to $^{11}$Be}

\begin{figure}[th]
\epsfxsize=7 cm
\centerline{\epsffile{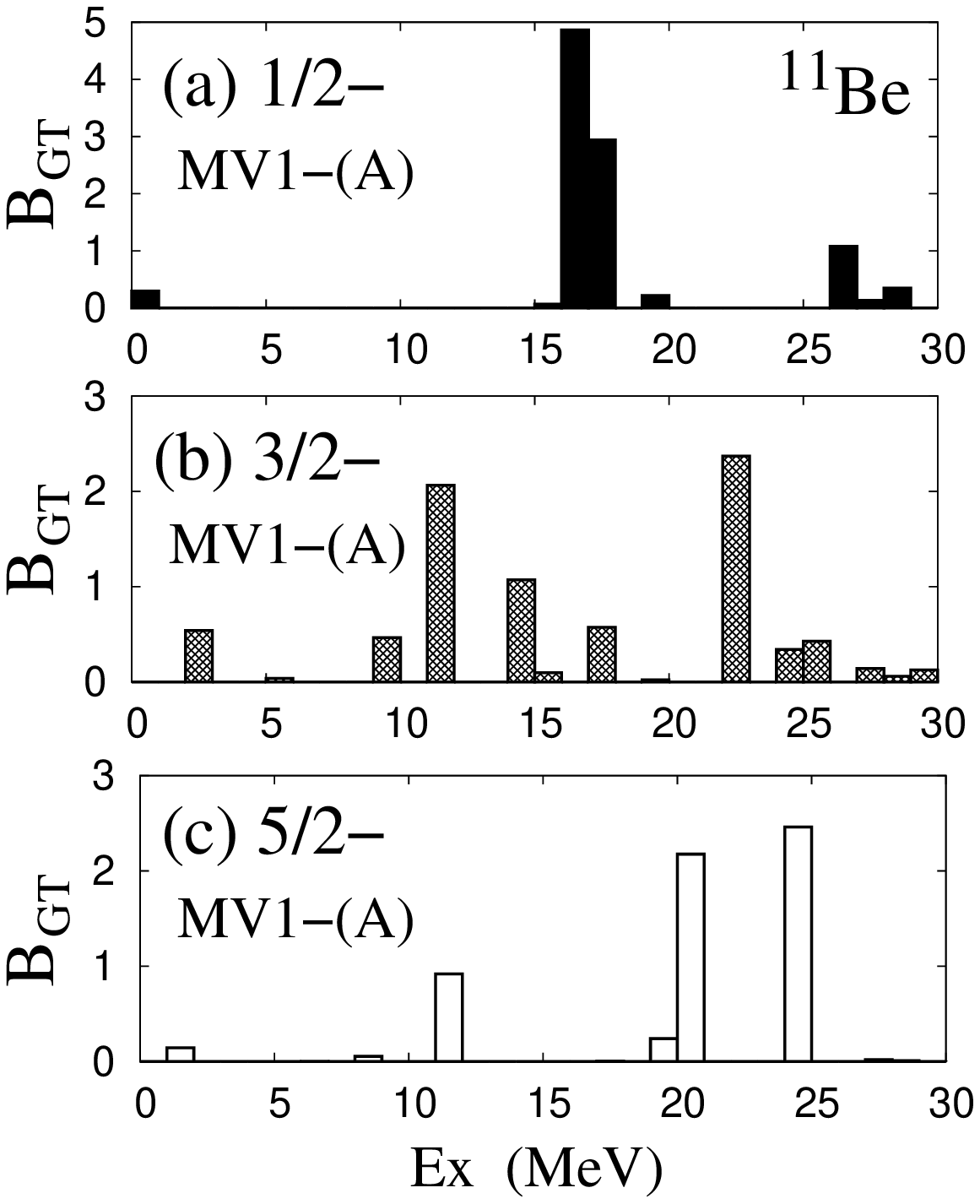}}
\vspace*{8pt}
\caption{
\label{fig:be11-gt}
The $B(GT)$ distribution to (a)$1/2^-$ states, 
(b)$3/2^-$ states and (c)$5/2^-$ states in the $^{11}$Li$\rightarrow^{11}$Be decay
calculated with the interaction set (A).
}
\end{figure}

\begin{figure}[th]
\epsfxsize=7 cm
\centerline{\epsffile{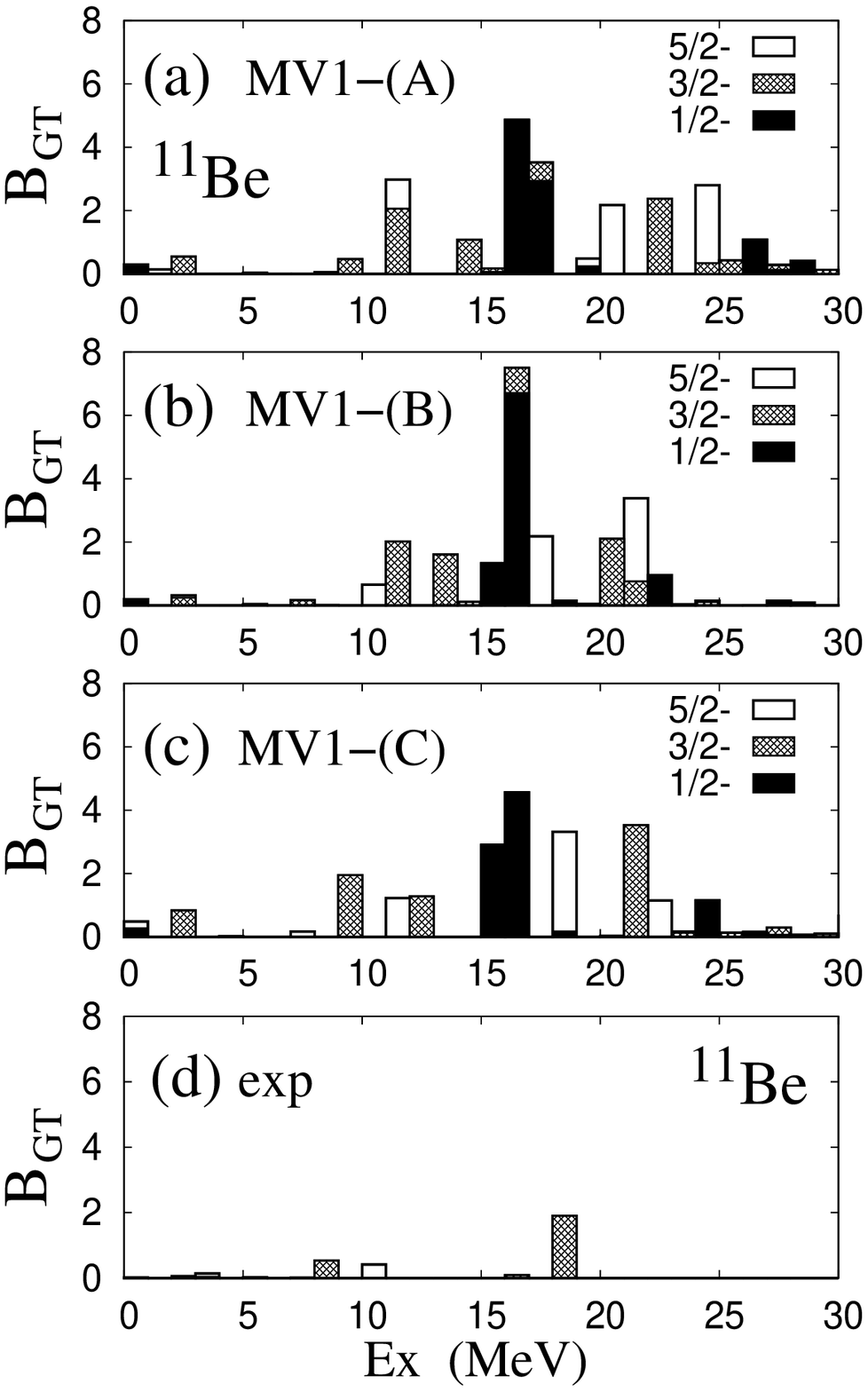}}
\vspace*{8pt}
\caption{
\label{fig:be11-gt-hist}
The $B(GT)$ distribution for the decay 
$^{11}$Li$\rightarrow ^{11}$Be. The $B(GT)$ values to $1/2^-$, $3/2^-$ and $5/2^-$ states
of $^{11}$Be are piled up.
The upper panels (a),(b) and (c) show 
the calculated values with the interaction sets (A), (B) and (C), respectively. 
The lowest panel (d) shows the experimental data 
taken from Refs.~\cite{Hirayama05,Madurga:2009ce}.}
\end{figure}

The $B(GT)$ distribution in the $^{11}$Li$\rightarrow^{11}$Be decay calculated with the interaction (A) 
is shown in Fig.~\ref{fig:be11-gt}. The $B(GT)$ values to $1/2^-$ states concentrate around $E_x=16-18$ MeV, and
those to $3/2^-$ states distribute widely in various excited states. 
The results of the $B(GT)$ distribution calculated with three sets (A), (B) and (C) of the interaction parameters
are displayed in Figs.~\ref{fig:be11-gt-hist}(a), \ref{fig:be11-gt-hist}(b) and \ref{fig:be11-gt-hist}(c) 
as well as the observed $B(GT)$ values
in Fig.~\ref{fig:be11-gt-hist}(d).
The results are qualitatively similar between these three interaction sets.
The $B(GT)$ values are small in $E_x < 10$ MeV, while they
distribute widely in $E_x=10-25$ MeV region and shows the broad peak structure 
with the center around $E_x=15-20$ MeV.
The small $B(GT)$ values to low-lying states are understood by the $2\alpha$ core
structure in $^{11}$Be, while 
the highly excited $^{11}$Be states with significant GT strengths are 
those with non-cluster or less-cluster structures. This is a similar situation to that in $^9$Be.


Although $^{11}$Li is known to be a neutron-rich nuclei with the neutron-halo structure,
halo effects are not taken into account in the present calculations because
single-particle wave functions in the present model are restricted to be a Gaussian form
and are not suitable to explain the long tail of the halo-neutron wave functions.
The halo effects in the $B(GT)$ values for the $^{11}$Li$\rightarrow^{11}$Be decay were discussed in the 
shell model study of Ref.~\cite{Suzuki:1997zza}, which showed that the halo structure gives 
small quenching of the $B(GT)$ values.

\subsection{Sum rule}

The Ikeda sum rule for the GT transition strengths is given as,
\begin{equation}
S^{(GT-)}-S^{(GT+)}=\frac{g^2_A}{g_V^2}\langle 0|[\mathbf{{\cal O}}^{(GT+)},
\mathbf{{\cal O}}^{(GT-)}]|0 \rangle\\
=\frac{g^2_A}{g_V^2}3(N-Z).
\end{equation}
Here  $S^{(GT\pm)}$ is the sum of $B(GT)$ values for the $\beta^\pm$ decay
and $|0 \rangle$ is the initial state.
Since $S^{(GT+)}$ for the $\beta^+$ decays of neutron-rich nuclei is very small, 
the sum rule is often approximated as
$S^{(GT-)}=\frac{g^2_A}{g_V^2}3(N-Z)$ by neglecting the $S^{(GT+)}$ term.

The values of the sum rule $3(N-Z)\times (g_A/g_V)^2$ are
13.59 and 22.65 for the $^9$Li and $^{11}$Li $\beta^-$ decays, respectively.
The theoretical values of sum of the $B(GT)$ values 
to $1/2^-$, $3/2^-$ and $5/2^-$ states in Be are listed in 
table~\ref{tab:bgt-sum}.
It is important that the sum values are determined only 
by the initial state of the Li ground state but do not depend on final states of Be.
Compared with the results (A), (B) and (C), it is found that 
interaction dependence of the sum values is small. It means that 
the wave functions of the $^9$Li and $^{11}$Li ground states are not sensitive to
the adopted interactions in the present calculations. 
The sum of the $B(GT)$ values to $1/2^-$ states is the largest. On the other hand, 
that to $5/2^-$ states is the smallest 
and exhausts only one forth of the calculated total sum $S^{(GT-)}$ value
in both the $^9$Li and $^{11}$Li decays. 
This result contradicts to the experimental report of the value
$B(GT)=8.9$ to the $5/2^-$ state at $E_x=11.81$ MeV in Ref.~\cite{Prezado03}.
Theoretically, the sum of the $B(GT)$ values for each spin is determined by the structure 
of the $^9$Li ground state with the spin and parity $3/2^-$, and it is unnatural that 
the highest-spin $5/2^-$ state exhausts such the large fraction of the Ikeda sum rule. 
In other words, 
the experimental value $B(GT)=8.9$ for the $5/2^-$ state seems to be 
too large to be described by theoretical calculations if 
the $^9$Li ground state has an ordinary structure.

\begin{table}[ht]
\caption{ 
\label{tab:bgt-sum} The calculated results for sum of the $B(GT)$ values
of $^9$Li and $^{11}$Li $\beta$ decays
to $J^-_f$ states in $^9$Be and $^{11}$Be. The values of Ikeda sum rule $3(N-Z)\times (g_A/g_V)^2$ are
13.59 and 22.65 for $^9$Li($\beta^-$) and $^{11}$Li($\beta^-$), respectively.}
\begin{center}
\begin{tabular}{c|cccc|cc}
\hline
 $^9$Li$(3/2^-_1)$ $\rightarrow$  $^9$Be$(J^\pi_f)$ & \multicolumn{3}{c}{$B(GT)$}\\
  $J^\pi_f$   & cal (A) & cal (B) & cal (C) \\
\hline
$1/2^-_1$ &6.9 &	7.0 	&7.1 \\
$3/2^-_1$ & 5.2 &	5.5 &	5.3 \\
$5/2^-_1$ & 3.6 &	3.2 &	3.5 \\
total & 15.7 &	15.7 &	15.9 \\
\hline
 $^{11}$Li$(3/2^-_1)$ $\rightarrow$  $^{11}$Be$(J^\pi_f)$ & \multicolumn{3}{c}{$B(GT)$}\\
  $J^\pi_f$   & cal (A) & cal (B) & cal (C) \\
\hline
$1/2^-_1$ & 10.0& 	9.5 &	10.1 \\
$3/2^-_1$ &8.5 &	8.1 	&8.5 \\
$5/2^-_1$ &6.2 &	5.7 	&6.2 \\
total& 24.7 &	23.3 	&24.7 \\
\hline
\end{tabular}
\end{center}
\end{table}

\section{Summary and outlook}\label{sec:summary}

The Gamow-Teller(GT) transitions in the $\beta^-$ decays, $^{9}$Li$\rightarrow ^{9}$Be
and $^{11}$Li$\rightarrow ^{11}$Be, were investigated 
with a method of antisymmetrized molecular dynamics.
The calculated $B(GT)$ values are small for transitions to low-lying states 
of Be isotopes because of the $2\alpha$-core structures in the final states.
Significant strengths are found in the $B(GT)$ distribution to 
non-cluster states of $^{9}$Be in $E_x\ge 10$ MeV region.
Sum of the $B(GT)$ values for each spin parity of the inal states was also 
studied.

Particular attention was paid to the strong $\beta$-transition from $^9$Li 
to the $^9$Be$(5/2^-)$ state at 11.8 MeV which was reported 
in the experimental work\cite{Prezado03}.
The present results are inconsistent with the strong GT transition to the
$5/2^-$ state which shows the large fraction of the Ikeda sum rule value.
Also in terms of the sum rule, 
the experimental value $B(GT)=8.9$ for the $5/2^-$ state seems to be 
too large to be described by theoretical calculations if 
the $^9$Li ground state has an ordinary structure.

This work is the first achievement in which the AMD method was applied to 
study of GT transitions to highly excited states.
One of the advantages of the present method is that
one can describe various final states in daughter nucleus(Be)
such as low-lying cluster states and high-lying non-cluster states.
Although the present model space is not a complete basis, it 
exhausts the GT transition strengths exactly. 

The present framework is a kind of bound state approximations and
continuum states are 
not incorporated. 
The coupling with continuum states should be taken into account carefully 
in discussion of widths of excited states above the thresholds.

\section*{Acknowledgments}
The computational calculations of this work were performed by using the
supercomputers at YITP and done in Supercomputer Projects 
of High Energy Accelerator Research Organization (KEK).
This work was supported by Grant-in-Aid for Scientific Research from Japan Society for the Promotion of Science (JSPS).
It was also supported by 
the Grant-in-Aid for the Global COE Program "The Next Generation of Physics, 
Spun from Universality and Emergence" from the Ministry of Education, Culture, Sports, Science and Technology (MEXT) of Japan.


\end{document}